\documentclass[preprint,showpacs,preprintnumbers,amsmath,amssymb]{revtex4}

\usepackage{graphicx}
\usepackage{dcolumn}
\usepackage{bm}
\usepackage{epsfig}

\def\k{{\bf k}} 
\def\mnras{MNRAS}
\def\HI{{\rm HI}}
\begin{document}

\title{Estimation of Cosmological Parameters from HI Observations of
 Post-reionization Epoch} 
 \author{Somnath
 Bharadwaj$^1$}\email{somnath@phys.iitkgp.ernet.in}\author{Shiv
 K. Sethi$^2$}\email{sethi@rri.res.in} \author{Tarun Deep
 Saini}\email{tarun@physics.iisc.ernet.in} 
\affiliation{${}^1$Department of Physics and Meteorology \& Center of
 Theoretical Studies,  
I.I.T. Kharagpur, 721302, India \\  ${}^2$Raman Research Institute,
 Bangalore 560080, India\\ ${}^3$Department of Physics, Indian
 Institute of Science, Bangalore 560 012} 

\begin{abstract}
  The emission from neutral hydrogen (HI) clouds in the
  post-reionization era ($z \le 6$), too faint to be individually
  detected, is present as a diffuse background in all low frequency
  radio observations below $1420 \, {\rm MHz}$.  The angular and
  frequency fluctuations of this radiation ($\sim 1 \, {\rm mK}$) is
  an important future probe of the large scale structures in the
  Universe. We show that such observations are a very effective probe
  of the background cosmological model and the perturbed Universe. In
  our study we focus on the possibility of determining the redshift
  space distortion parameter $\beta$, coordinate distance $r_{\nu}$,
  and its derivative with redshift, $r'_{\nu}$. Using reasonable
  estimates for the observational uncertainties and configurations
  representative of the ongoing and upcoming radio interferometers, we
  predict parameter estimation at a precision comparable with
  supernova Ia observations and galaxy redshift surveys, across a wide
  range in redshift that is only partially accessed by other probes.
  Future HI observations of the post-reionization era present a new
  technique, complementing several existing one, to probe the
  expansion history and to elucidate the nature of the dark energy.
\end{abstract}


\maketitle

\section{Introduction}
Determining the expansion history of our Universe and parameterizing
the constituents of the Universe at a high level of precision, are
currently some of the most important goals in cosmology. While
high-redshift ($z \leq 2$) supernova~Ia observations (e.g.
\cite{riess,perlmutter}) and galaxy surveys ($z \leq 1$ ) (e.g.
\cite{tegmark}) probe the local universe; and CMBR observations (e.g.
\cite{dunkley,komatsu}) probe the recombination era $(z \sim 1000)$,
the expansion history is largely unconstrained across the vast
intervening redshift range.  Observations of redshifted $21 \,{\rm
  cm}$ radiation from neutral hydrogen (HI) hold the potential of
probing the universe over a large redshift range ($20 \ge z \ge 0$):
from the dark ages to to the present epoch (eg.  \cite{BA5,furla}).
Such observations can possibly be realized at several redshifts, using
the currently functioning GMRT
\footnote{http://www.gmrt.ncra.tifr.res.in/}. Several new telescopes
are currently being built with such observations in mind (eg. MWA
\footnote{http://www.haystack.mit.edu/ast/arrays/mwa/} \& LOFAR
\footnote{http://www.lofar.org/}). Such observations will map out the
large-scale HI distribution at high redshifts. It has recently been
proposed \cite{wlg,chang} that Baryon Acoustic Oscillations (BAO) in
the redshifted $21 \,{\rm cm}$ signal from the post-reionization era
($z \le 6$) is a very sensitive probe of the dark energy. The BAO is a
relatively small ($\sim 10-15$ per cent) feature that sits on the HI
large-scale structure (LSS) power spectrum. In this paper we
investigate the possibility of probing the expansion history in the
post-reionization era using the HI LSS power spectrum without
reference to the BAO. Unless otherwise stated we use the parameters
$(\Omega_{m0},\Omega_{\Lambda0}, \Omega_b h^2, h,
n_s,\sigma_8)=(0.3,0.7, 0.024,0.7,1.0,1.0)$ referred to as the LCDM
model in our analysis.

At redshifts $z \le 6$, the bulk of the neutral gas is in clouds that
have HI column densities in excess of $2 \times 10^{20}\,\,{\rm
  atoms/cm^{2}}$ \cite{peroux,lombardi,lanzetta}. These high column
density clouds are observed as damped Lyman-$\alpha$ absorption lines
seen in quasar spectra. These observations indicate that the ratio of
the density $\rho_{\rm gas}(z)$ of neutral gas to the present critical
density $\rho_{\rm crit}$, of the universe has a nearly constant value
$\rho_{\rm gas}(z)/\rho_{\rm crit} \sim 10^{-3}$, over a large
redshift range $0 \le z \le 3.5$. This implies that the mean neutral
fraction of the hydrogen gas is $ \bar{x}_{\HI}=50\,\,\Omega_{\rm gas}
h^2 (0.02/\Omega_b h^2) =2.45 \times 10^{-2}$, which we adopt for the
entire redshift range $z \le 6$.  The redshifted $21 \, {\rm cm}$
radiation from the HI in this redshift range will be seen in emission.
The emission from individual clouds ($ < 10 \,\mu{\rm Jy}$) is too
weak to be detected with existing instruments unless the image is
significantly magnified by gravitational lensing \cite{saini}. The
collective emission from the undetected clouds appears as a very faint
background in all radio observations at frequencies below $1420 \,
{\rm MHz}$. The fluctuations in this background with angle and
frequency is a direct probe of the HI distribution at the redshift $z$
where the radiation originated. It is possible to probe the HI power
spectrum at high redshifts by quantifying the the fluctuations in this
radiation (\cite{bns,bs}).

\section{Formulation}

The Multi-frequency Angular Power Spectrum ( MAPS) 
$C_{\ell}(\Delta \nu)$ \cite{datta1}  quantifies the statistics of the
HI signal 
 as a joint function of the angular multipole $\ell$ and the
frequency separation $\Delta \nu$. We
define  the angular power spectrum 
$C_{\ell}=C_{\ell}(0)$ and  the frequency decorrelation function 
\begin{equation}
\kappa_{\ell}(\Delta \nu)=\frac{C_{\ell}(\Delta \nu)}{C_{\ell}(0)}\,\,, 
\label{eq:kappa}
\end{equation}
to separately characterize the angular and the $\Delta \nu$ dependence
respectively. The latter quantifies whether the HI signal at two
different frequencies $\nu$ and $\nu+\Delta \nu$ is correlated
$\kappa_{\ell}(\Delta \nu) \sim 1$ or uncorrelated
$\kappa_{\ell}(\Delta \nu) \sim 0$ . The function $C_{\ell}(\Delta
\nu)$ can be estimated directly from observations without reference to
a cosmological model (eg. \cite{ali08}).  However, it is necessary to
assume a background cosmological model in order to interpret
$C_{\ell}(\Delta \nu)$ in terms of the three dimensional LSS HI power
spectrum.  On the large scales of interest here, it is reasonable to
assume that HI traces the dark matter with a possible linear bias $b$,
whereby the three dimensional HI power spectrum is $b^2 P(k)$, where
$P(k)$ is the dark matter power spectrum at the redshift where HI
signal originated. We have \cite{datta1}
\begin{equation}
C_l(\Delta \nu)=
\frac{\bar{T}^2~ }{\pi r_{\nu}^2}
\int_{0}^{\infty} {\rm d} k_{\parallel} \, 
\cos (k_{\parallel}\, r'_{\nu}\, \Delta \nu) \, P_{\rm HI}({\bf k}) \,,
\label{eq:fsa} 
\end{equation}
where the three dimensional wavevector ${\bf k}$ has been decomposed
into components $k_{\parallel}$ and $l/r_{\nu}$, along the line of
sight and in the plane of the sky respectively. The comoving distance
$r_{\nu}$ is the distance at which the HI radiation originated. Note
that $(1+z)^{-1} \, r_{\nu}=d_{\rm A}(z)$ is the angular diameter
distance and $r_{\nu}^{'}=d r_{\nu}/d \nu$. The temperature occurring
in eq.~(\ref{eq:fsa}) is given by
\begin{equation}
\bar{T}(z)=4.0 \, {\rm mK}\,\,(1+z)^2  \, \left(\frac{\Omega_b
  h^2}{0.02}\right)  \left(\frac{0.7}{h} \right) \frac{H_0}{H(z)} \,,
\label{eq:a5}
\end{equation}
and $P_{\rm HI}({\bf k})$ is the three dimensional power spectrum of
the ``21 cm radiation efficiency in redshift space'', which in this
situation is given by
\begin{equation}
P_{\rm HI}(\k)=\bar{x}^2_{\HI} b^2 \left( 1+ \beta  \mu^2 \right)^2
P(k) \,.
\label{eq:d1}
\end{equation}
The term $\left( 1+ \beta \mu^2 \right)^2$ arises due to HI peculiar
velocities (\cite{bns,bharad04}), which we assume to be determined by
the dark matter. This is the familiar redshift space distortion seen
in galaxy redshift surveys, where $\mu=k_{\parallel}/k$. and $\beta
=f(z)/b$ is the linear distortion parameter, which is the ratio of
$f(z)$ that quantifies the growth rate of linear perturbations, and
$b$ the linear bias.

\section{Results and Conclusions}
\begin{figure}
\begin{center}
\mbox{\epsfig{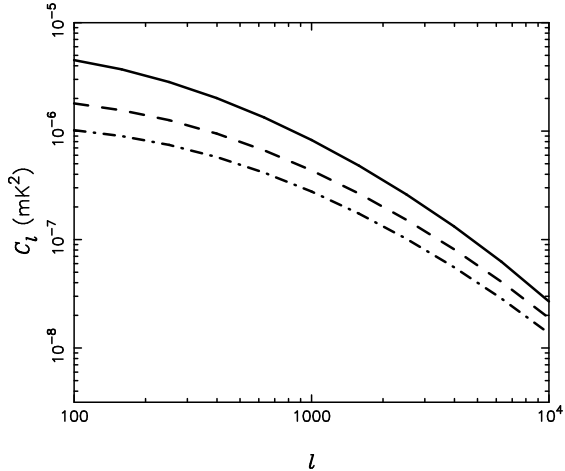}}
\caption{Here we plot $C_l(0)$  at redshifts $z=\{1.5,3.0,4.5\}$. 
  The signal decreases monotonically with increasing redshift, so the
  lowest plot is for the highest redshift. We assume the bias to be
  $b=1$ throughout.  }
\label{fig:cl}
\end{center}

\end{figure}
\begin{figure}
\begin{center}
  \mbox{\epsfig{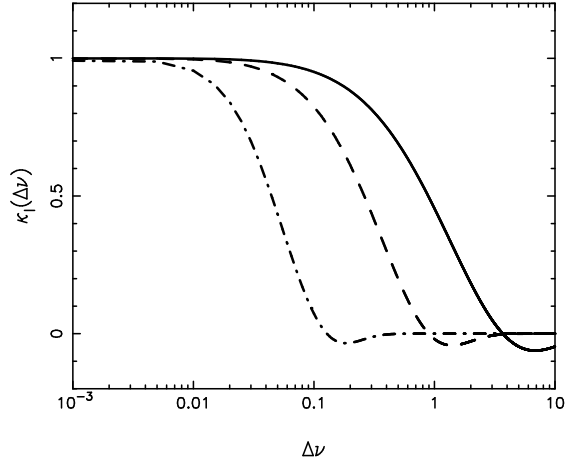}}
\caption{Here we plot the frequency decorrelation function $\kappa_{\ell}(\Delta \nu)$ 
  as a function of $\Delta \nu$, for a fixed redshift $z=3.0$ and
  $\ell=\{100,1000,10000\}$. The signal declines more sharply
  for higher value of $\ell$.}
\label{fig:kappa}
\end{center}
\end{figure}

The expected signal $C_l(\Delta \nu)$ from a few representative
redshifts, calculated for the LCDM model, is plotted in
Figure~\ref{fig:cl}, and in Figure~\ref{fig:kappa} we have plotted the
frequency decorrelation function $\kappa_{\ell}(\Delta \nu)$ as a
function of $\Delta \nu$, for a fixed redshift $z=3.0$ and for
$\ell=100,\,1000 \,\& \,10000$. The HI signal is smaller than $\sim 1
\, {\rm mK}$, and it decreases with increasing $l$.  The shape or
$\ell$ dependence is decided by the shape of $P(k)$ at all comoving
wave-numbers $k \ge \ell/r_{\nu}$.  The signal at two different
frequencies $\nu$ and $\nu+\Delta \nu$ decorrelates rapidly with
increasing $\Delta \nu$ and $\kappa_{\ell}(\Delta \nu) < 0.1$ at
$\Delta \nu > 5 \, {\rm MHz}$.  The decorrelation occurs at a smaller
$\Delta \nu$ for the larger multipoles (Figure \ref{fig:kappa}). While
the HI signal at a frequency separation $\Delta \nu>5\, {\rm MHz}$ is
expected to be uncorrelated, the foregrounds are expected to be highly
correlated even at frequency separations larger than this (eg.
\cite{santos}).  This should in principle allow the HI signal to be
separated from the foregrounds, which are a few orders of magnitude
larger (eg. \cite{mcquinn,mor2}).

\begin{figure}
\begin{center}
\mbox{\epsfig{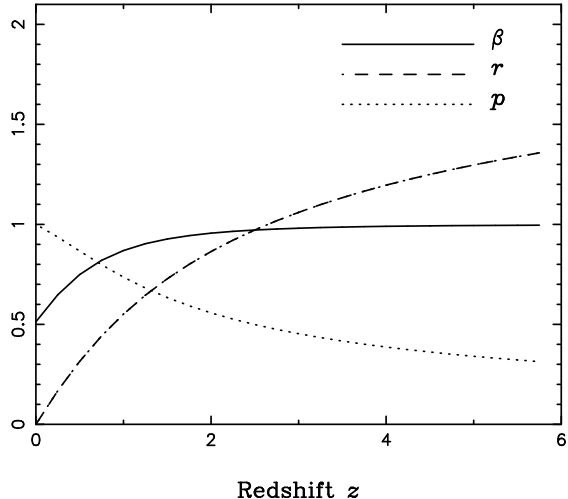}}
\end{center}
\caption{Here we plot the parameters $(\beta,r,p)$ as a function
  of redshift $z$ for the concordance LCDM model. The parameter $r=r_{\nu}/(6000\,\,{\rm Mpc)}$.}
\label{fig:parm}
\end{figure}

It is clear from eq.~(\ref{eq:fsa}) that $C_{\ell}(\Delta \nu)$
depends on the background cosmological model through the parameters
$(\beta,r_{\nu},r^{'}_{\nu})$.  Assuming that the dark matter power
spectrum $P(k)$ is known a priori, observations of
$C_{\ell}(\Delta\nu)$ can be used to determine the values of these
three parameters. It is convenient to replace $r^{'}_{\nu}$ with the
dimensionless parameter \cite{ali1}
\begin{equation}
p(z)=\frac{d \ln\left [r_{\nu}(z) \right]}{d
  \ln(z)} \,.
\end{equation}
Figure~\ref{fig:parm} shows the variation of the three parameters
$(\beta,r_{\nu},p)$ across the redshift range $z\le 6$ for the LCDM
model.

We separately consider parameter estimation using $C_{\ell}$ and
$\kappa_{\ell}(\Delta \nu)$. The former does not depend on $p$. The
amplitude $A= (\bar{T} \bar{x}_{\HI} b)^2 /\pi r_{\nu}^2$ of
$C_{\ell}$ is uncertain, and we consider the joint estimation of three
parameters $(A,\beta,r_{\nu})$ from observations of $C_{\ell}$.  The
value of $\kappa_{\ell}(\Delta \nu)$ is insensitive to the amplitude
$A$, leaving three parameters $(\beta,r_{\nu},p)$ that can be jointly
estimated from this.  We use the Fisher matrix (e.g. \cite{tth}) to
determine the accuracy at which these parameters can be estimated.

Parameter estimation depends on two distinct aspects of the observing
instrument. The first is the $\ell$ range {\it ie.}  $\ell_{min}$,
$\ell_{max}$, and the sampling interval $\Delta \ell$, which
corresponds to the smallest $\ell$ spacing at which we have
independent estimates of $C_{\ell}(\Delta \nu)$ . This is determined
by the instrument's field of view, and is inversely related to it.
The second is the observational uncertainty in $C_{\ell}(\Delta \nu)$.
This is a sum, in quadrature, of the instrumental noise and the cosmic
variance.  The cosmic variance contribution $\delta
C_{\ell}/C_{\ell}=\sqrt{{2}/{((2 \ell +1) \, {\rm f} \, \Delta
    \ell})}$ (${\rm f}$ is the fraction of sky observed) is further
reduced because the large frequency bandwidth $\Delta\nu_B$ provides
several independent estimates of $C_{\ell}$. We assume that $\delta
C_{\ell}$ is reduced by a factor we $\sqrt{\Delta\nu_B/(1 \, {\rm
    MHz})}$ because of this. The instrumental uncertainties were
estimated using relations \cite{ali08} between $\delta C_{\ell}$ and
the noise in the individual visibilities measured in
radio-interferometric observations.  For this we assume that the
baselines in the radio-interferometric array have a uniform u-v
coverage.

We consider three different instrumental configurations  for
parameter estimation.

\begin{itemize}
\item[A.] The currently functional GMRT has too few antennas for
  cosmological parameter estimation. We consider an enhanced version
  of the GMRT with a substantially larger number of antennas ($N=120$)
  , each identical to those of the existing GMRT. The antennas have a
  relatively small field of view ($\theta_{\rm FWHM}\sim 0.8^{\circ}$
  at $610 \, {\rm MHz}$) and the array has relatively large baselines
  spanning $\ell_{min}=500$ to $\ell_{max}=10,000$ with $\Delta
  \ell=100$.
\item[B.] The upcoming MWA will have a large number of small sized
  antennas.  The antennas have a relatively large field of view
  ($\theta_{\rm FWHM}\sim 5^{\circ}$ at $610 \, {\rm MHz}$), and the
  array is expected to be quite compact spanning $\ell_{min}=100$ to
  $\ell_{max}=2000$ with $\Delta \ell=20$.  The first version of this
  array is expected to have $N=500$ antennas which is what we
  consider.
\item[C.] This is a future, upgraded version of the MWA which is
  expected to have $N=5000$ antennas.
\end{itemize}
For each of these configurations, we assume that 16 simultaneous
primary beams can be observed.  We present results for $2$ years of
observation for A and B, and $1000$ hours for C.  Throughout we assume
frequency channels $0.05 \, {\rm MHz}$ wide, a bandwidth $\Delta \nu_B
=32 \, {\rm MHz}$, and that a single field is observed for the entire
duration. For parameter estimation we use: $\delta
\kappa_{\ell}(\Delta \nu)=\sqrt{2} \, \delta C_{\ell}/C_{\ell}$.

\begin{figure*}
\begin{center}
\mbox{\epsfig{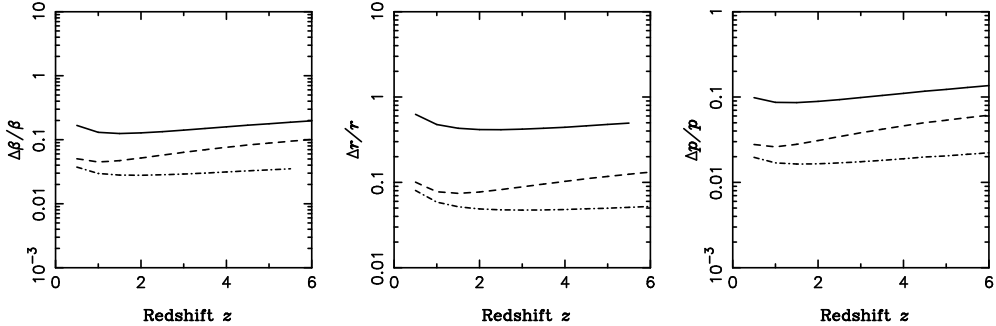}}
\end{center} 
\caption{Expected one-sigma fractional errors for parameter estimation at
  different redshifts for the LCDM model. The curves in each panel
correspond, from top to bottom, to the cases A, B, and C, respectively. }
\label{fig:par}
\end{figure*}

\begin{figure}
\begin{center}
\mbox{\epsfig{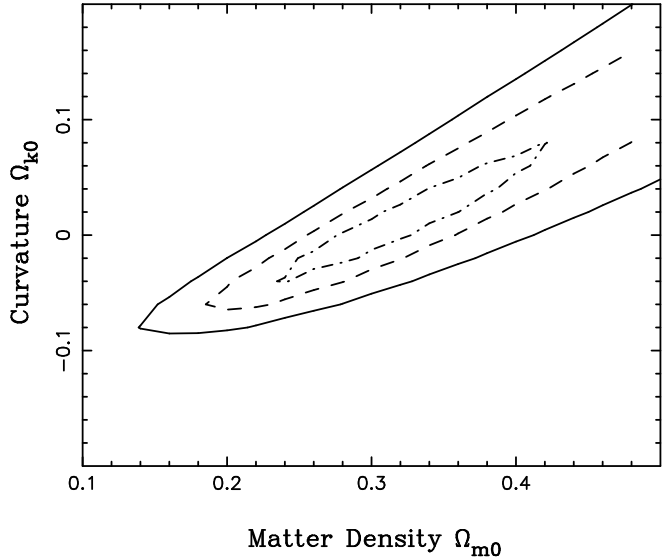}}
\end{center}
\caption{Expected  one-sigma confidence regions for the
  parameters $\Omega_{m0}$ and $\Omega_{k0}$, based on estimated
  errors for observations of $p$, corresponding to Figure~4, at $z=3$. }
\label{fig:contour}
\end{figure}

We find that observations of $C_{\ell}$ impose very poor constraints
on the parameters $\beta$ and $r_{\nu}$, and we do not show these
here.  The accuracy is considerable higher for $\kappa_{\ell}(\Delta
\nu)$, which captures the three dimensional clustering of the HI as
compared to $C_{\ell}$, which quantifies only the angular dependence.
Figure~\ref{fig:par} shows the predicted estimates for the parameters
$\beta$, $r_{\nu}$ and $p$ at various redshifts.  Further, we find
that a compact, wide-field array (B,C) is considerably more sensitive
to these parameter as compared to case A.

Considering the three parameters individually:

{\bf Redshift-space distortion parameter: $\beta$ }. This has
traditionally been measured from galaxy redshift surveys
\cite{Peacock,Hawkins,Ross,Guzzo}, with uncertainties in the range
$0.1 \le \Delta \beta/\beta \le 0.2$. These observations have, till
date, been restricted to $z \le 1$. Future galaxy surveys are expected
to achieve higher redshifts and smaller uncertainties. Galaxy surveys
have the drawback that at very high redshifts they probe only the most
luminous objects, which are expected to be highly biased. HI
observations do not have this limitation and could provide high
precision $(\Delta \beta/\beta <0.1)$ estimates over a large redshift
range.

{\bf Coordinate distance, $r_\nu$}: The most direct measurement of the
coordinate distance comes from supernova type~Ia observations for $z
\le 2$.  Current Sn~Ia observations give $\Delta r_\nu/r_\nu \simeq
0.07$ \citep{sai04} for a single supernova. The statistical error in
the coordinate distance can be further reduced by observing a large
number of supernovae in a small redshift bin; thus the fundamental
limitation of this technique is due to unknown systematics in the
supernovae themselves, since it is certainly possible that supernovae
at high redshift are different. Figure~4 shows that the HI method
might have the potential to enable a precise measurement of the
coordinate distance up to much larger redshifts.  Furthermore, such a
complimentary probe will also help in ascertaining systematics in the
supernova probe.

{\bf Derivative of coordinate distance, p}: This quantifies the
Alcock-Paczynski (AP) effect \cite{Alcock}, which is well accepted as
a means to study the expansion history at high $z$, though such
observations have not been possible till date.  Observations of
redshifted $21\, {\rm cm}$ radiation hold the potential of measuring
the AP effect \cite{Nusser,ali1,Barkana}.  The parameter $p$ is not
affected by the overall amplitude $A$ and the bias $b$, and is a
sensitive probe of the spatial curvature (Figure \ref{fig:parm}).  Our
estimates indicate that it will be possible to measure $p$ with an
accuracy $\Delta p/p \sim 0.03$ over a large $z$ range.

The parameters $(\beta,r_{\nu}, p)$ chosen for our analysis occur
naturally when we interpret $C_{\ell}(\Delta \nu)$ in terms of the
three dimensional dark matter power spectrum $P(k)$. Further, these
parameters are very general in that they do not refer to any specific
model for either the dark energy or the dark matter, and are valid
even in models with alternate theories of gravity (eg.
\cite{Carroll,Dvali}). In fact, observations of these three parameters
at different redshifts can in principle be used to distinguish between
these possibilities.

For the purpose of this paper, we illustrate the cosmological
parameter estimation by considering the simplest LCDM model, with two
unknown parameters $\Omega_{m0}$ and $\Omega_{k0}$, and
$\Omega_{\Lambda0}=1-\Omega_{m0}-\Omega_{k0}$. In
Figure~\ref{fig:contour} we plot the $1\hbox{--}\sigma$ confidence
interval for the estimation of $\Omega_{m0}$ and $\Omega_{k0}$, using
a single measurement of $p$ alone, {\it ie.} only one of the three
parameters measured at a single redshift $z=3$.  Note that $p$ is
insensitive to $H_0$ and hence it is not considered as an additional
parameter here.  It is possible to combine measurements at different
$z$ to improve the constraints on cosmological parameters.  We shall
undertake a detailed analysis for quantifying the precision that can
be achieved by combining different data sets (CMBR, galaxy surveys)
for a more complicated dark energy model in a future work.

In conclusion, HI observations of the post-reionization era can, in
principle, determine the expansion history at a high level of
precision and thereby constrain cosmological models.  Neither the
upcoming initial version of the MWA which is planned to have $ 500$
antenna elements nor any conceivable upgradation of the existing GMRT
will be in a position to carry out such observations, the observation
time needed being too large.  We find that an enhanced version of the
MWA, which is planned to have $5000$ antenna elements, would be in a
position to meaningfully constrain cosmological models.  By combining
different probes, we expect to achieve an unprecedented precision in
the determination of cosmological parameters. This will be a step
towards pinning down the precise nature of dark energy in the
universe.

\bibliography{apssamp}

\end{document}